\documentstyle[sprocl,epsf]{article}

\def\Journal#1#2#3#4{{#1} {\bf #2}, #3 (#4)}

\def\NIMA{{\em Nucl. Instrum. Methods} A}

\def\PRL{\em Phys. Rev. Lett.}

\def\EPJ{{\em Eur. Phys. J.}}

\def\ks{K^0_s}

\def\be{\begin{equation}}
\def\ee{\end{equation}}
\def\bea{\begin{eqnarray}}
\def\eea{\end{eqnarray}}

\def\B0{B_d^0} 
\def\B0b{\overline{B}{^0}{_d}} 
\def\BJKS{B_d^0/\overline{B}{^0}{_d} \rightarrow J/\psi K^0_s} 
 
\def\B0JKS{B_d^0 \rightarrow J/\psi K^0_s} 
\def\JKS{J/\psi K^0_s} 
\def\jpsi{J/\psi} 

\begin{document}


\title{FLAVOR TAGGING AND CP-VIOLATION MEASUREMENTS AT THE TEVATRON}

\author{S. TKACZYK \\(representing the CDF Collaboration)}

\address{Fermi National Accelerator Laboratory, P. O. Box 500, 
Batavia, IL 60510, USA\\E-mail: tka@fnal.gov} 


\maketitle\abstracts{ The CDF collaboration has adapted several
heavy flavor tagging techniques and employed them in analyses
of time-dependent flavor asymmetries using data from the Tevatron
Run I. The tagging algorithms were calibrated using low-$P_t$ 
inclusive lepton and dilepton trigger data samples.  
The tagging techniques were applied to a sample of $\sim 400$ 
$\BJKS$ decays and were used to measure the CP violation parameter,
{$\sin(2\beta$)}. Prospects for future improved measurements
of the CP violation parameters at the Tevatron are briefly discussed.}

\section{Introduction}

CP violation was first observed in kaon decays over 30 years ago.
In the Standard Model the Cabibbo-Kobayashi-Maskawa (CKM) weak 
and mass eigenstates mixing matrix can provide a possible mechanism
for explanation of the observed CP violation effects.
The unitary CKM matrix is described by four physical parameters,
one of them being a complex phase. 

An analysis of unitarity constraints in which 
all of the elements are of the same order of magnitude,
e.g.:
 $V_{ud}V_{ub}^* + V_{cd}V_{cb}^* + V_{td}V_{tb}^* = 0$ 
and
 $V_{ud}V_{td}^* + V_{us}V_{ts}^* + V_{ub}V_{tb}^* = 0$ 
\noindent
provides a rudimentary test of the CKM description of CP violation. 
The magnitude of the complex elements have been determined
from B-hadron lifetimes, branching fractions and 
- more recently - precise flavor oscillations measurements.
The relative complex phases of the CKM matrix elements can be studied
in measurements of the CP asymmetries in B-decays.  
An analysis of the asymmetry in the decay rates of $B^0$ and $\bar{B}^0$
to a common CP eigenstate $\JKS$
provides a measurement of the 
phase $\beta \equiv arg(- \frac{ V_{cd}V_{cb}^*}{V_{td}V_{tb}^*} )$.
The asymmetry, ${\cal{A}}_{CP} \equiv \frac{N(\bar{B}^0) - N(B^0)}
{N(\bar{B}^0) + N(B^0)}$, where $N(\bar{B}^0)$ and $N(B^0)$ 
are numbers of observed decays to $\JKS$  given the known flavor of 
the B meson at production, arises from the interference between
the direct decay path, 
 $\bar{B^0} \rightarrow J/\psi K^0_s$,  and the mixed decay path,
 $\bar{B^0} \rightarrow B^0 \rightarrow J/\psi K^0_s$. 

The CP asymmetry ${\cal{A}}_{CP}$ depends on the CP phase difference 
between the two amplitudes, $\beta$ 
and the flavor oscillations term represented
by  $\sin(\Delta m_d t)$, where $\Delta m_d$ is the mass difference 
between the two $B_d^0$ mass eigenstates, and {\it t} is proper decay time. 
In the Standard Model ${\cal{A}}_{CP} \simeq \sin(2\beta) \sin(\Delta m_d t)$ 
since other contributions are expected to be very small.
Values of $\sin(2\beta)$ are 
constrained to a range of $0.3 \leq  \sin(2\beta) \leq 0.9$ from 
indirect electroweak measurements.\cite{Mele} Last year
the OPAL collaboration at LEP reported \cite{Opal}
$\sin(2\beta)=4\pm2\pm1$, using
a sample of 14 $B^0/\bar{B^0} \rightarrow J/\psi K^0_s$ decays.

In this talk I will describe the flavor tagging techniques adapted
by  CDF  for application in the hadron collider
environment and discuss their performance 
in the flavor oscillation measurements.
I will also report on the CP analysis of 
$\B0JKS$ decays reconstructed in a data sample of 
110 pb$^{-1}$ collected by the CDF detector at the Tevatron collider 
at Fermilab. The description of the CDF detector 
can be found in previous publications \cite{cdfd1,cdfd2}.   

\section{Data Sample}

The reconstruction of $B^0_d$ mesons was done via the 
decay $\BJKS$, with $J/\psi \rightarrow \mu^+ \mu^-$ and 
$K^0_s \rightarrow \pi^+ \pi^-$.
The selection of the $B$ candidates begins by identifying $\jpsi$
particles that decay into two muons of opposite charge. 
\begin{figure}[ht]
 \centerline{
  \epsfysize 6.0cm
  \epsffile{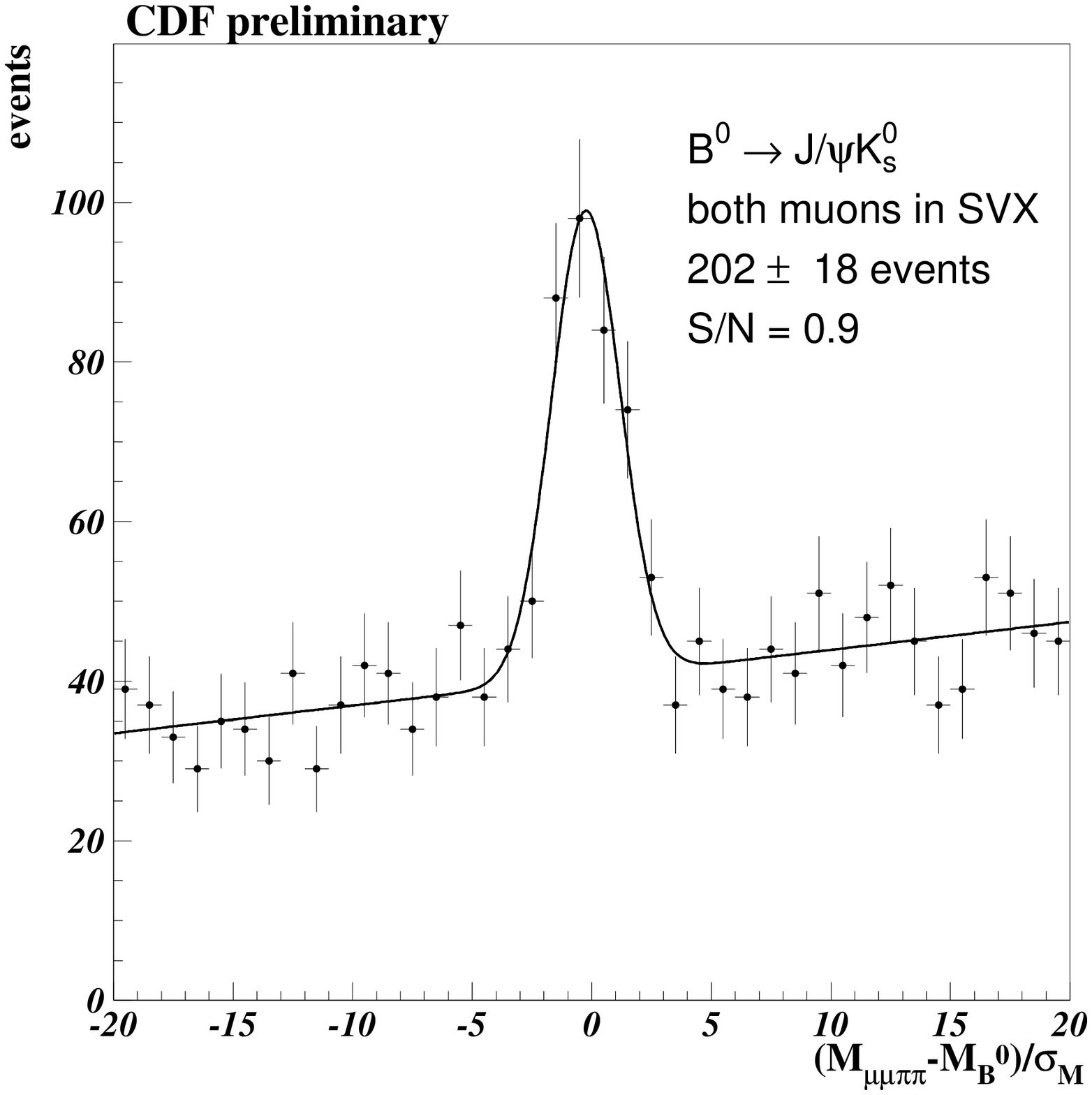}
  \epsfysize 6.0cm
 \epsffile{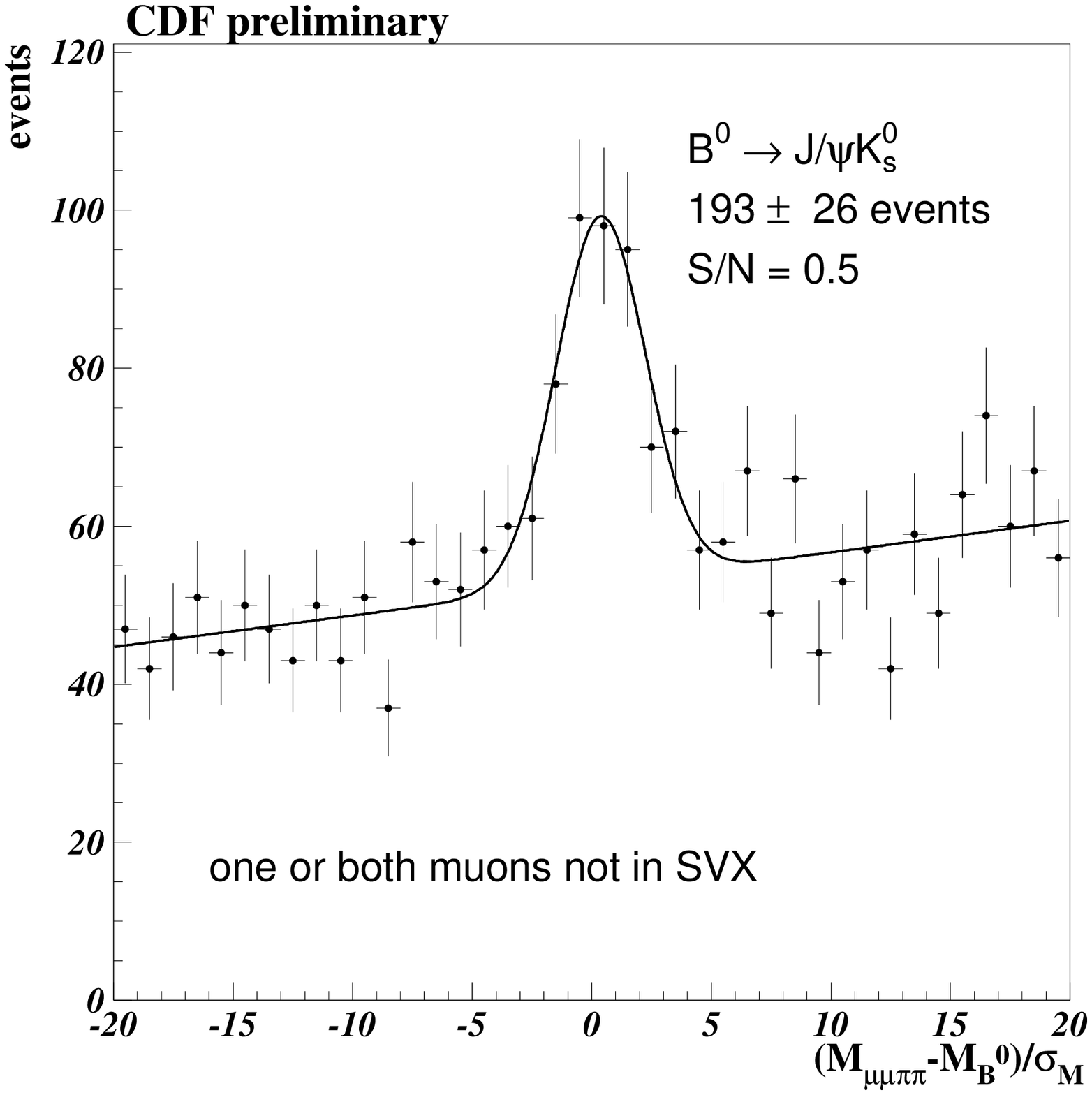}}
\caption{Normalized mass distributions of the $\B0JKS$  candidates.
 \label{fig:jpsiks}}
\end{figure}
\noindent
All pairs of the oppositely charged particle tracks are considered to be
candidates for the $\ks$ decay products. 
The B candidate  mass and momentum are calculated subject to the 
constraints that the invariant masses of the muon 
pair and the pion pair are equal to the world average mass of their parent
particle, $\jpsi$  and $\ks$, respectively; 
come from separate vertices; the reconstructed $\ks$ 
candidate points back  to the $\jpsi$ vertex; and the  $\JKS$ system 
points back to the primary  interaction  vertex. 
The silicon micro-vertex
detector (SVX) information was used for these constraints when available.
For a B candidate with both muons measured in the silicon vertex detector,
the typical mass resolution is $\sim 10~MeV/c^2$, 
and the proper decay length  resolution is $\sim 50~\mu m$.  
The  normalized mass distribution, 
$M_N = (m_{\mu \mu \pi \pi} - M_{PDG})/\sigma_{fit}$, 
is shown in figure ~\ref{fig:jpsiks}. The total number
of reconstructed B mesons is $395\pm31$, with a signal-to-noise 
ratio of 0.7. The sample with both muons reconstructed in SVX contains
202$\pm$18 events with a signal-to-noise ratio of 0.9, and the remainder 
of the sample contains 193$\pm$26 events with a signal-to-noise ratio of 0.5.   

\vspace*{-0.2cm}
\section{Identification of B Flavor at Production and Decay} 

\subsection{Opposite Side Tagging with Soft Lepton and Jet Charge}\label{subsec:slt}

\begin{figure}[ht]
 \centerline{
  \epsfysize 6.0cm
  \epsffile{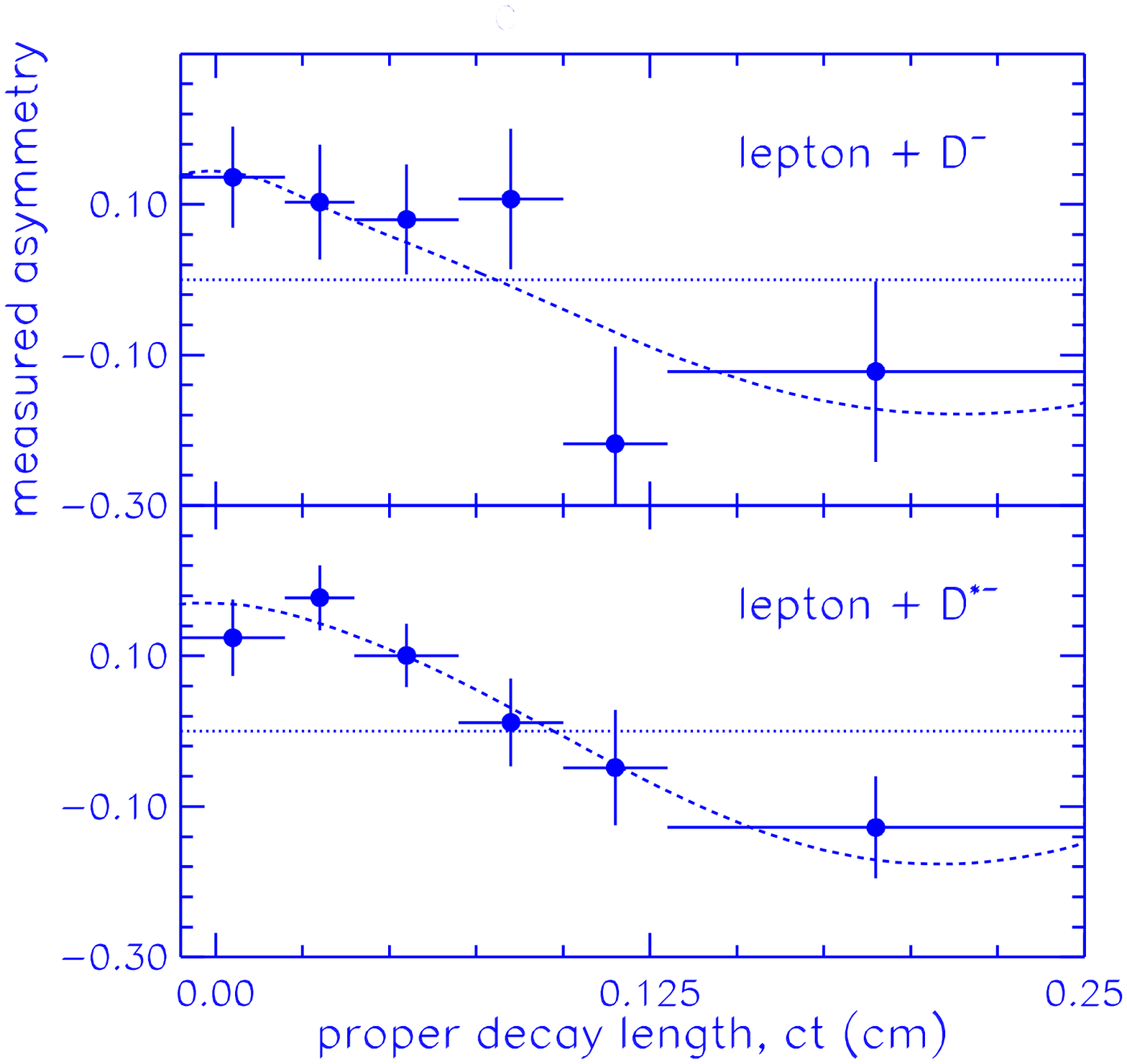}
  \epsfysize 6.0cm
  \epsffile{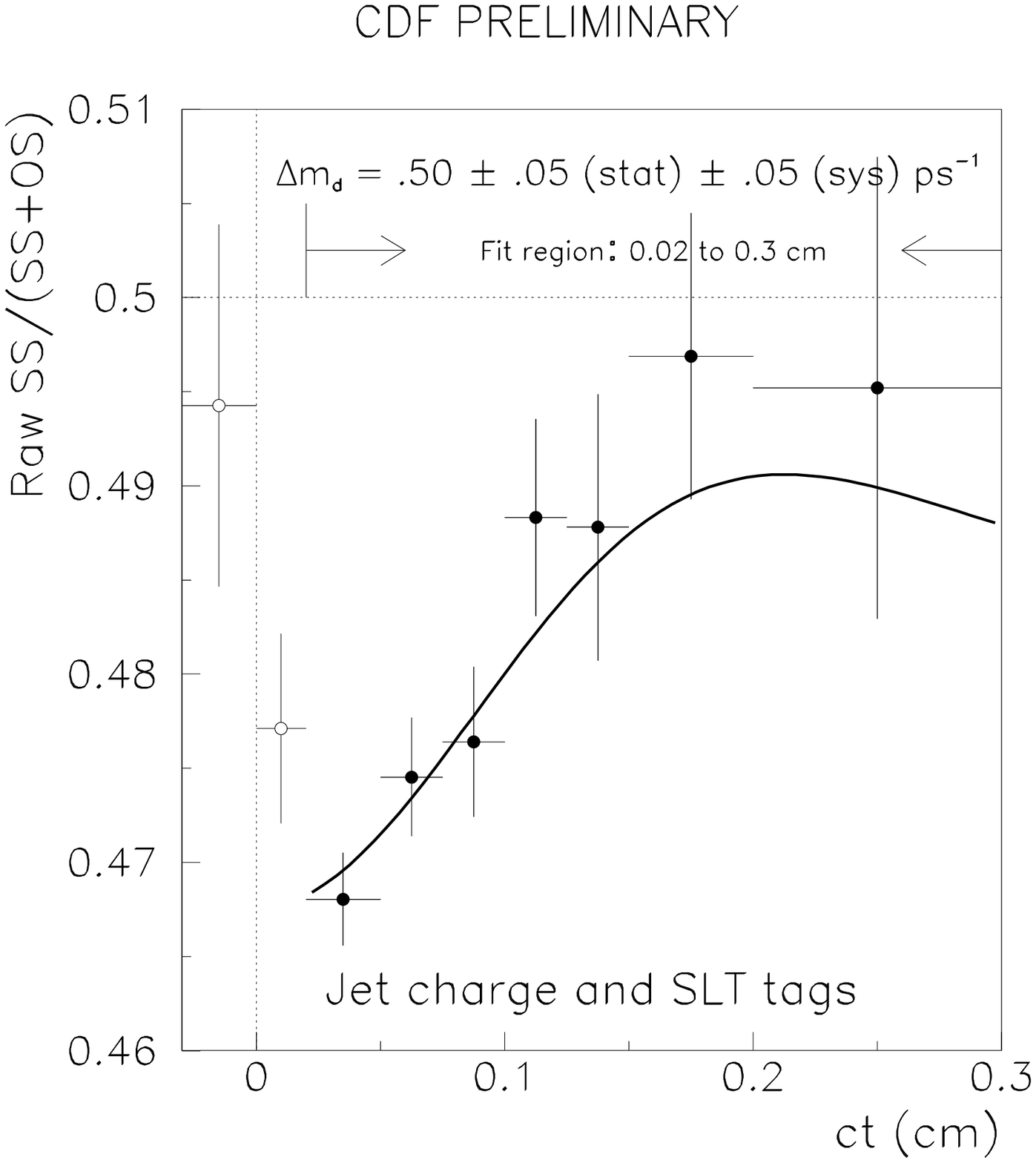}}
\caption{Asymmetry as a function of the proper decay length ct.
Left: Same Side tagging applied to $B \rightarrow \ell D^{*}$;
Right: Soft lepton and Jet Charge flavor tagging. 
Results from an unbinned likelihood fit are
 superimposed on the data points.  \label{fig:bbar}}
\end{figure}

 The Opposite Side Flavor Tagging techniques
use a triggered lepton  and a reconstructed secondary vertex to identify
the flavor of the B meson at the decay time. The flavor of the B 
meson at the production time is determined either from the charge
of the jet on the side opposite the triggered lepton or by
the presence of another lepton in the event.    
These flavor tagging methods were studied using high statistics samples
 of semileptonic B decays,\cite{cdfsltjqc,lepd} as illustrated in 
figure ~\ref{fig:bbar}.
The Opposite Side Flavor tagging algorithms were calibrated using a sample
of  $\sim 1,000$ $B_u^{\pm} \rightarrow J/\psi K^{\pm}$ decays.

\vspace{-0.3cm}

The Soft Lepton Tagging (SLT) algorithm correlates the charge of 
the second lepton in the event with the flavor of the B at the 
production time.
Its performance was checked through observation of the 
$B_d^0-\overline{B_d^0}$ flavor oscillation using an inclusive 
lepton trigger sample,\cite{cdfsltjqc} as  shown in fig.   
~\ref{fig:bbar}. The dilution of the soft lepton tag, as
measured using the $J/\psi K^+$ sample, is ${\cal D}$
 = 63 $\pm$ 15 \%.

In the Jet Charge (JTQ) method a momentum weighted charge average of
 particles
 in a b-jet, $Q_{jet}$, is used to determine the charge of the b quark.   
The event is considered as tagged when $|Q_{jet}|>0.2$.
The performance of the  JTQ  was also checked with the 
analysis of the $\Delta m$ and dilution ${\cal D}$
 using the inclusive lepton trigger sample (fig.   
~\ref{fig:bbar}).
 The dilution of the JTQ method, as measured with the
 $J/\psi K^+$ sample, is ${\cal D}$ = 24 $\pm$ 7\%.
A summary of the performance of tagging algorithms, described by the 
value of the dilution and the tagging efficiency,\footnote{
The tagging efficiency, $\epsilon$, is the fraction of 
 events that are tagged.} is presented in table~\ref{tab:eps}. 
The Jet Charge and Soft Lepton tagging algorithms are described in more
detail in another CDF publication.\cite{cdfsltjqc}

\subsection{Same Side Tagging}\label{subsec:sst}

The Same Side Tagging (SST) technique relies on the correlation
between the flavor of the B hadron and the charge of a nearby 
hadron produced either in the fragmentation process that
formed a B meson from a $b$ quark or from the decay of $B^{**}$ meson.
The charge correlations are the same in both cases:
a $B^0_d$ meson is associated with a positive particle.
The SST algorithm selects as a flavor tag, that particle which has 
the minimum momentum component transverse  to the momentum 
sum of the B and the particle. The particle has to
 be contained in an $\eta - \phi$ 
cone of 0.7 around the B momentum direction, have $P_t > 400~ MeV/c$,
and come from the primary vertex.    
The performance of this method was calibrated by tagging $B \rightarrow 
 \ell D^{(*)}$ decays and observing the time dependence  of the 
$B^0_d \overline{B^0_d}$ oscillation,\cite{lepd}
 as shown in fig. \ref{fig:bbar}.
 In addition to the usual measurement of the frequency of the 
oscillation $\Delta m_d$, the amplitude of the oscillation, $\cal{D}$,
called dilution,\footnote{The tagging dilution is defined as 
${\cal{D}} = \frac{N_R - N_W}{N_R + N_W}$, where $N_W(N_R)$ are numbers of
wrong (right) tagging decisions. The observed asymmetry
${\cal A}_{CP}^{obs}={\cal D} {\cal A}_{CP}$.} was also determined. 

\noindent

\section{Flavor Asymmetry in $\B0JKS$ Sample}

Following the conclusion of the Workshop, the CDF collaboration 
updated the previously published analysis \cite{cdfsin2b} of the
$\B0JKS$ decays by 
employing a combination of three tagging methods to the full sample
of $\sim$ 400 events. The SST and SLT algorithms were essentially
 the same 
 as in those used in the $\ell D^{(*)}$\cite{lepd}
 and inclusive lepton\cite{cdfsltjqc} analyses.
 The JTQ  algorithm
was  modified  from that in ref. 5
to increase the efficiency of identifying 
low-$P_t$ jets. Each event can be tagged by one algorithm on the opposite 
side and one on the same side. When both SLT and JTQ tags are
available, the tagger with higher dilution was selected 
to avoid introduction of correlations. In addition, the dilution and
efficiencies for the opposite side tagging algorithms were determined 
using a sample of $B \rightarrow J/\psi K^+$ decays, to match the 
kinematic properties of the two samples.

\vspace{-0.5cm}

\begin{figure}[ht]
 \centerline{
  \epsfysize 6.5cm
\epsffile{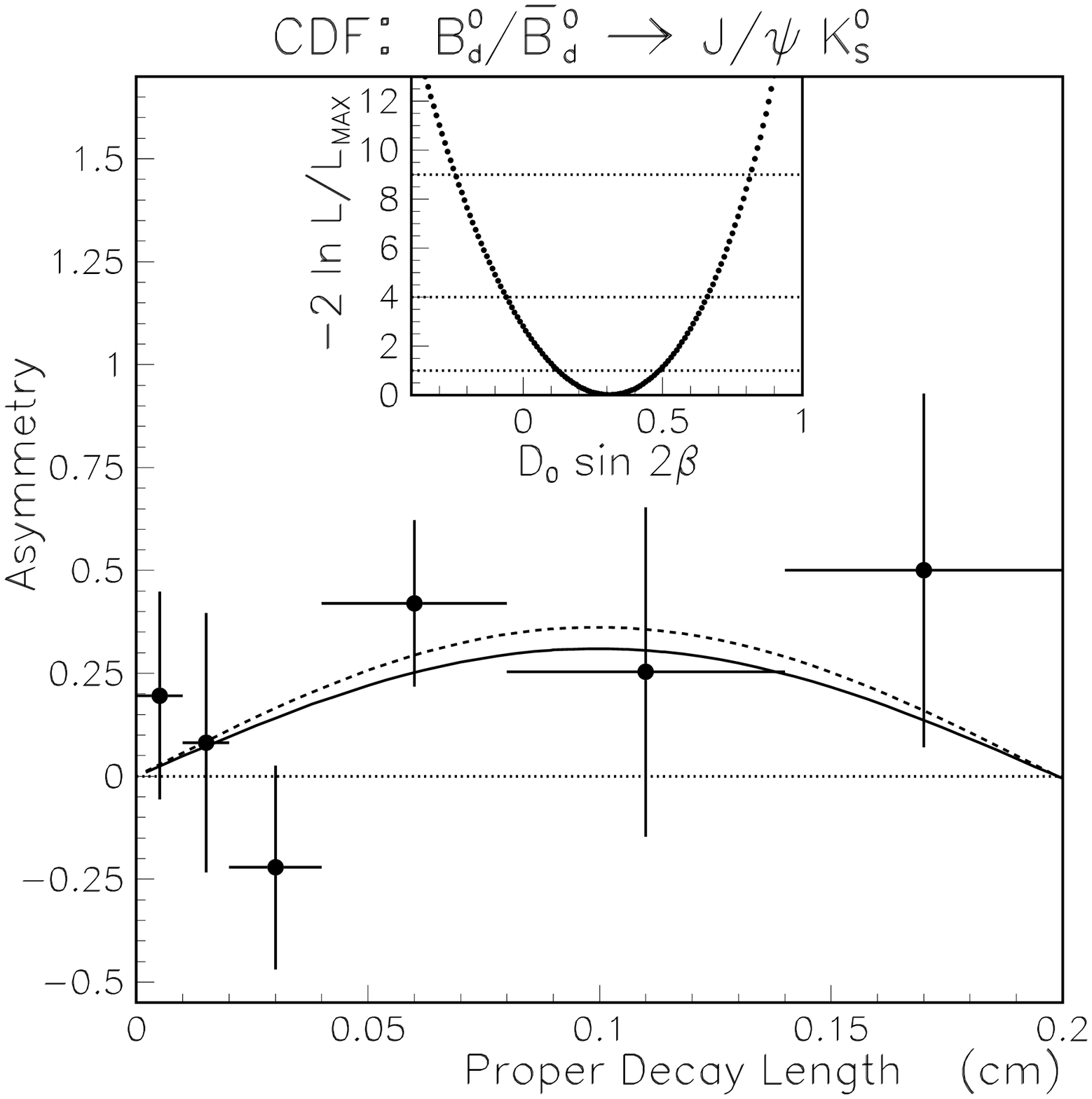}
  \epsfysize 6.5cm
\epsffile{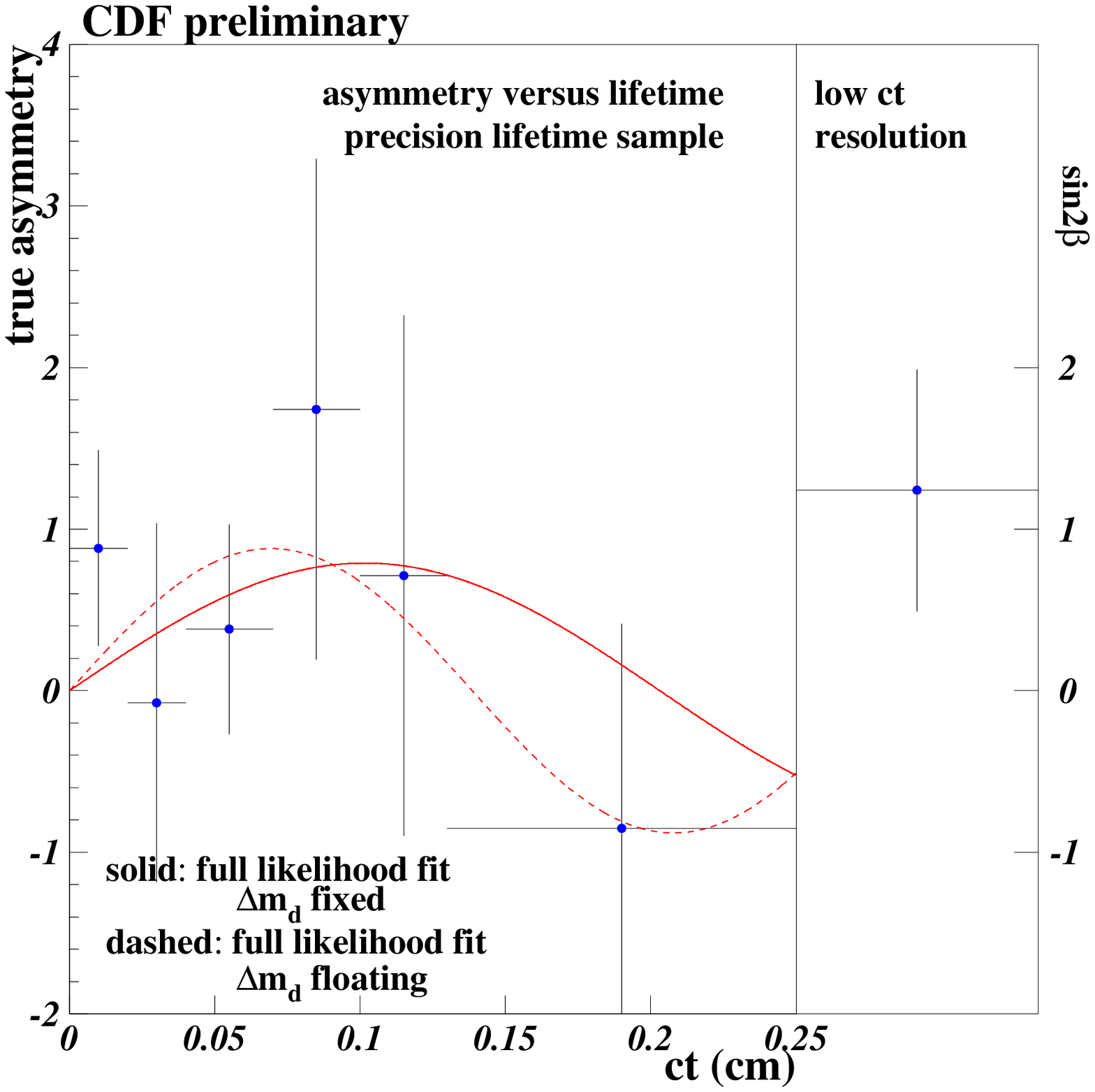}}
\caption{Results of CP asymmetry studies. Left: Time dependent 
Same Side Tag applied to a sample of $\B0JKS$, where two muons are
reconstructed in the SVX  detector. Right: Multiple tagging analysis results.
In addition to the time dependent information the plot displays 
time-integrated asymmetry for non-SVX events.  For comparison
the dashed line present results of the fit with $\Delta m_d$
left floating in the fit.
\label{fig:sin2beta}}
\end{figure}

Tagged events are simultaneously fitted for a combination of the three 
tagging methods, using an unbinned likelihood fit with the value 
of  $\Delta m_d$ fixed to the world value. 
 The fitting  also takes into account the remaining tag correlations.
The asymmetry values for the three tagging methods are shown in 
fig. \ref{fig:sin2beta}. Those events
 without proper time determination are presented separately as 
a single point.
We measure   $\sin(2\beta ) = 0.79 ^{+0.41}_{-0.44}$.
The curves shown in fig.~\ref{fig:sin2beta} present the results of the 
fit.

\begin{table}[t]
\caption{Summary of the statistical power of the taggers, measured by 
$\epsilon D^2$.\label{tab:eps}}
\vspace{0.2cm}
\begin{center}
\footnotesize
\begin{tabular}{|c|c|c|c|}
\hline
Tagger & \raisebox{0pt}[13pt][7pt]{Effective Dilution} &
\raisebox{0pt}[13pt][7pt]{Dilution} &{Efficiency}\\
 & \raisebox{0pt}[13pt][7pt]{$\epsilon {\cal D}^2$} &
\raisebox{0pt}[13pt][7pt]{${\cal D}$} &{$\epsilon$}\\
\hline
SST   & $2.1 \pm 0.5$ \% &  17 $\pm$ 4\% & 38$\pm$ 4\% \\
JTQ   & $2.2 \pm 1.3$ \% &  24 $\pm$ 7\% & 40$\pm$ 4\% \\
SLT   & $2.2 \pm 1.0$ \% &  63 $\pm$ 15\% & 6 $\pm$ 2\% \\
\hline
Multiple Tags   & $6.3 \pm 1.7$  \% &   & \\
\hline
\end{tabular}
\end{center}
\end{table}

\section{Conclusion}

Multiple tagging methods have been validated in the hadron collider 
environment of the Tevatron. The statistical power of the taggers,
 measured by the quantity $\epsilon {\cal D}^2$, was determined using data 
sets accumulated by the CDF collaboration.
 Using a sample of over $\sim 400$ events of fully reconstructed $\BJKS$
decays and multiple tags, we measured  $\sin(2\beta ) =
 0.79 ^{+0.41}_{-0.44}$.
This result can be translated into the frequentist confidence interval
of $0 < \sin(2\beta) <1$ at 93\% confidence level.
Next year, the Tevatron will begin a new collider run, delivering 
an expected twenty-fold  increase in data over the following two years.
With detector and trigger improvements, we expect to accumulate a sample of
 $\sim$ 10,000 $\BJKS$ events,
allowing the uncertainty on  $\sin(2\beta )$ to be reduced to 0.08.    
All three tagging methods  will be important tools in the study of 
CP violation effects in the next run of the Tevatron.

\section*{Acknowledgments}

     We thank the Fermilab staff and the technical staffs of the
participating institutions for their vital contributions.  This work was
supported by the U.S. Department of Energy and National Science Foundation;
the Italian Istituto Nazionale di Fisica Nucleare; the Ministry of Education,
Science and Culture of Japan; the Natural Sciences and Engineering Research
Council of Canada; the National Science Council of the Republic of China; 
the Swiss National Science Foundation; the A. P. Sloan Foundation; and the
Bundesministerium fuer Bildung und Forschung, Germany.

\end{document}